\documentclass[aps,pra,showpacs,showkeys,superscriptaddress,10pt,preprintnumbers,reprint]{revtex4-1}

\usepackage[utf8]{inputenc}	
\usepackage[T1]{fontenc}
\usepackage[english]{babel}

\usepackage{amsmath}
\usepackage{amsfonts}
\usepackage{latexsym}
\usepackage{amssymb} 
\usepackage{MnSymbol} 
\usepackage{bbm} 
\usepackage{textcomp}

\usepackage[pdftex]{graphicx} 

\usepackage[usenames]{color}

\newcommand{\bref}[1]{(\ref{#1})}

\newcommand{\br}{{\bf R}}

\newcommand{\bS}{{\bf \hat S}}
\newcommand{\itext}[1]{\text{\it #1}}

\newcommand{\hP}{\mathcal{\hat P}}

\DeclareSymbolFont{bbold}{U}{bbold}{m}{n}
\DeclareSymbolFontAlphabet{\mathbbold}{bbold}

\usepackage{hyperref}

\begin{document}

\preprint{\scriptsize \it It's an ArXiv copy of the paper published in Acta Physica Polonica A {\bf 126}, A-25 (2014),
DOI:10.12693/APhysPolA.126.A-25}

\title{$t$--$t'$--$J$--$U$ model in mean-field approximation: coexistence of superconductivity and antiferromagnetism}

\author{Marcin Abram}
\email{marcin.abram@uj.edu.pl}
\affiliation{Marian Smoluchowski Institute of Physics, Jagiellonian University, Reymonta 4, 30-059 Krak\'ow, Poland}

\date{\today}


\begin{abstract}
We discuss the $t$--$J$--$U$ model in the mean-field approximation. The role of spin-exchange coupling $J$ and the second nearest hopping $t'$ are examined
in the context of the coexistence of superconductivity (SC) and antiferromagnetism (AF). Stability of the phases is studied with respect to temperature.
The coexistence region exists for the sufficiently large Coulomb repulsion ($U>U_{cr}$), and in the vicinity of the half-filled band (hole doping $\delta < \delta_{cr}$).
The critical hole doping is relatively small ($\delta_{cr} \approx 0.006$ for $J/|t| = 1/3$) and linear with respect to $J$. The decrease of $U_{cr}$ is proportional to~$J$,
except the limit of small~$J$ ($J/|t|< 0.03)$, where $U_{cr}$ grows rapidly with decreasing~$J$.
The effect of the second nearest hopping is limited -- the phase diagram does not change in a~qualitative manner when the $t'$ value is changed.
In the limit of $T \rightarrow 0$, SC phase is stable even for large hole-doping (such as $\delta = 0.5$).
Additional paramagnetic (PM) phase appears for large $\delta$ or small $U$ at non-zero temperature. When temperature increases, both SC and AF+SC phase regions
are reduced.
\end{abstract}



\maketitle


\section{Intorduction}

One of the basic models for high-temperature superconductors and correlated systems is $t$--$J$ model, which can be derived from the Hubbard model
in the limit of large Coulomb repulsion $U$ \cite{Spatek1977-PhysicaBC.86-88.375, *ChaoSpalekOles1977-JPhysC.10.L271, Spalek2007-ActaPhysPolonA.111.409}.
In the simplest version 
the $t$--$J$ model has the form \cite{Spatek1977-PhysicaBC.86-88.375, *ChaoSpalekOles1977-JPhysC.10.L271, Spalek2007-ActaPhysPolonA.111.409, Dagotto1994-RevModPhys.66.763, Lee2006-RevModPhys.78.17}
\begin{equation}
\begin{split}
   \mathcal{\hat H}_{t-J} = & \sum_{i \neq j,\, \sigma} \hP_0\, t_{ij} \hat c_{i\sigma}^\dagger \hat c_{j \sigma}\, \hP_0 \\
& + \sum_{i \neq j} J_{ij}\, \hP_0 \left( \bS_i \cdot \bS_{j} - \frac{1}{4} \hat n_i \hat n_j \right) \hP_0 , \label{eq:H_tJ}
\end{split}
\end{equation}
where $t_{ij}$ is the hopping integral, $J_{ij} \equiv 4 t_{ij}^2/U$ is the kinetic-exchange integral, and $\hP_0 = \prod_i (1- \hat n_{i\uparrow} \hat n_{i\downarrow})$
is the Gutzwiller projector operator eliminating the double site occupancies. Sometimes, for simplicity, the term $\frac{1}{4} \hat n_i \hat n_j$ is neglected
(cf.\ discussion of the term's relevance in Ref.~\onlinecite{Jedrak2011-PhD}, Chap.~9).

For Hubbard model, the energy cost for two electrons residing on the same site is equal to~$U$, hence in the limit of $U \rightarrow +\infty$
(which was assumed when deriving the $t$--$J$ model \cite{Spatek1977-PhysicaBC.86-88.375}), the double occupancies are prohibited. It is realized through the projector
$\hP_0$ which eliminates them. Alternatively, interaction term of the Hubbard type, $U \sum_i \hat n_{i\uparrow} \hat n_{i\downarrow}$, can be added
to the Hamiltonian (\ref{eq:H_tJ}) explicitly. In such situation and for sufficiently large $U$, the energy of the double occupancies is high so that they effectively
are not present in the system. In effect, the projector $\hP_0$ can be omitted (cf.\ e.g.\ Ref.~\onlinecite{Lin1991-PRB.44.4674}, where such approach was formulated). 

However, one could argue, that e.g.\ for the cuprates, the term proportional to $J_{ij}$ does not only reflect the kinetic exchange interactions of $d$-holes in the Cu plane,
but also incorporates effects of the Cu-O hybridization, hence the $J_{ij} \equiv 4 t_{ij}^2/U$ identity is no longer valid \cite{Zhang1988-PhysRevB.37.3759}.
Furthermore, the Cu-O hybridization can reduce the cost of  double occupancy, and the requirement of large $U$ may no longer be necessary. Thus, the enlarged Hamiltonian becomes effective
and all three parameters, $t_{ij}$, $J_{ij}$, and $U$, can now be treated as independent parameters.
This can be regarded, as rationale for introducing the $t$--$J$--$U$ model. 

The $t$--$J$--$U$ model was extensively studied by Zhang \cite{Zhang2003-prl}, Gan \cite{Gan2005-prb, Gan2005-prl}, and Bernevig \cite{Bernevig-PhysRevLett.91.147003}.
However, no antiferromagnetic order was considered in those works
\footnote{Some attempts was made by some authors, cf.\ PRB {\bf 71}, 104505 (2005), PRA {\bf 79} 063611 (2009), and J. Phys.: Condens Matter {\bf 23}, 495602 (2011),
but their methods suffered from some serious inconsistencies which affected their final results (cf.\ discussion in Ref.~\protect\onlinecite{Abram2013-PRB.88.094502}).}.
Recently, we have covered the topic (cf.\ Ref.~\onlinecite{Abram2013-PRB.88.094502}) and we have found that in the $t$--$J$--$U$ model for sufficiently large~$U$,
a~coexistence of antiferromagnetism and superconductivity (AF+SC) appears, but only in a very limited hole-doping (close range to the half-filled band).
The present article is an extension of the previous work~\cite{Abram2013-PRB.88.094502}. The model is refined to consider also the second nearest-neighbor hopping.

The structure of this paper is as follows: in Sec.~\ref{sec:model} the model is defined, as well as the approximations leading to the effective single-particle Hamiltonian.
In Sec.~\ref{sec:SGA} the details of the solving procedure are provided. Results and discussions are presented in Secs.~\ref{sec:Results} and \ref{sec:Conclusions}, respectively.


\section{The model and the effective single-particle Hamiltonian}
\label{sec:model}

The starting Hamiltonian for $t$--$J$--$U$ model has the form \cite{Zhang2003-prl, Gan2005-prb, Gan2005-prl}
\begin{equation}
   \mathcal{\hat H} = \sum_{i \neq j,\, \sigma} t_{ij} \hat c_{i\sigma}^\dagger \hat c_{j \sigma} + \sum_{i \neq j} J_{ij}\, \bS_i \cdot \bS_{j} + U \sum_i \hat n_{i\uparrow} \hat n_{i\downarrow}, \label{eq:H_tJU}
\end{equation}
where $t_{ij}$ denotes the hopping term, $J_{ij}$ the spin-exchange coupling, $U$ the on-site Coulomb repulsion, $\hat c_{i\sigma}^\dagger$~($\hat c_{i\sigma}$) are
creation (annihilation) operators of an electron on site $i$ and with spin $\sigma$; $\hat n_{i\sigma} \equiv  \hat c_{i\sigma}^\dagger \hat c_{i \sigma}$ denotes
electron number operator, $\bS_i \equiv (\hat S_i^x ,\, \hat S_i^y,\, \hat S_i^z)$ spin operator. In the fermion representation 
$\hat S_i^\sigma \equiv \frac{1}{2} (\hat S_i^x + \sigma \hat S_i^y) = \hat c_{i\sigma}^\dagger \hat c_{i\sigma}$, while  $\hat S_i^z = \frac{1}{2}(\hat n_{i\uparrow} - \hat n_{i\downarrow})$.

\begin{figure}
 \centering
 \includegraphics[width=0.44\columnwidth]{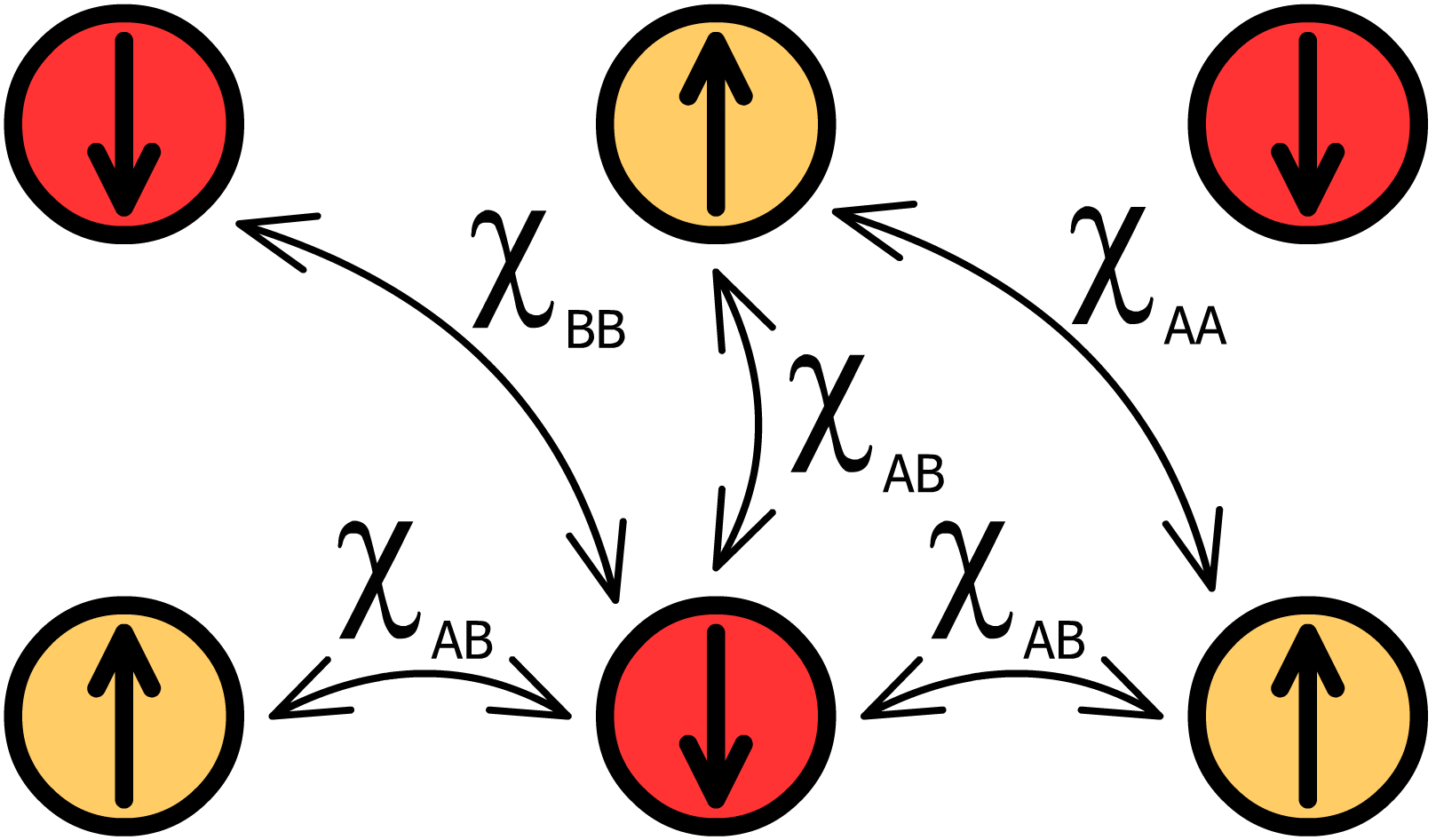}\hspace{12pt}
 \hspace{-8pt} \vline \hspace{8pt} 
 \includegraphics[width=0.44 \columnwidth]{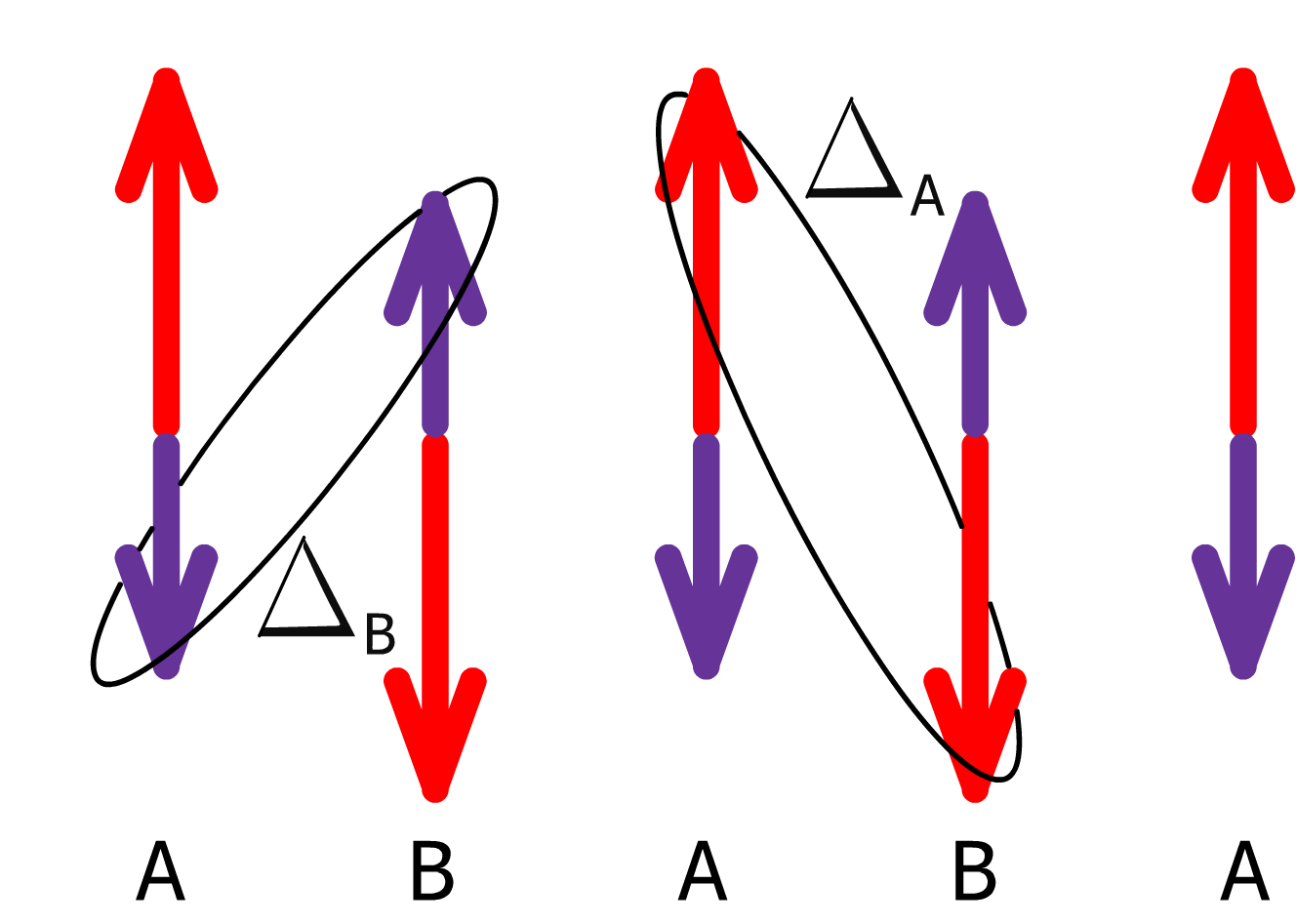}
 \caption{(Color online) Schematic interpretation of $\chi$, $\chi_{AA} $ and $\chi_{BB}$ (\underline{left} panel) and $\Delta_A$ and $\Delta_B$ (\underline{right} panel). To consider antiferromagnetism in the system, we can divide the lattice into two sublattices, $A$ where in average the spin is \emph{up}, and $B$ where in average is \emph{down}. Thus $\chi$ denotes hopping between sites belonging to sublattices $A$ and $B$, while $\chi_{AA}$ and $\chi_{BB}$ hopping within one sublattice ($A$ or $B$ respectively); $\Delta_A$ denotes pairing of majority spins (\emph{up} from sublattice $A$ and \emph{down} form $B$), and $\Delta_B$ pairing of minority spins (\emph{up} from $B$ and \emph{down} from $A$).}\label{sites}
\end{figure}

Here, we consider a~two-dimensional, square lattice. This is justified since cuprates have a~quasi two-dimensional structure. We assume that $J_{ij} \equiv J/2$ if $i$, $j$ indicate
the nearest neighbors, and $J_{ij} \equal 0$ otherwise. We restrict hopping to the first ($t$) and the second nearest neighbors ($t'$).
We use the Gutzwiller approach (GA) \cite{Gutzwiller1963-prl, Gutzwiller1965-pr} to obtain an effective single-particle Hamiltonian. Specifically, to calculate the average
$\langle \mathcal{\mathcal{\hat H}} \rangle \equiv \langle \Psi \mid  \mathcal{\mathcal{\hat H}} \mid  \Psi \rangle$, the form of $|\Psi \rangle$ has to be known.
We are assuming that $|\Psi \rangle \approx |\Psi_G \rangle \equiv \hat P_G |\Psi_0\rangle= \prod_i \big( 1-(1-g ) \hat n_{i\uparrow} \hat n_{i\downarrow} \big) |\Psi_0\rangle$,
where $g$ is a~variational parameter and $|\Psi_0\rangle$ is a~single-particle wave function. Note, that for $g=0$ the projector cuts off all states with double occupation
(two electrons on one site), while for $g=1$ we have simple $| \Psi_G \rangle = | \Psi_0 \rangle$. 
In GA, we assume that
\begin{equation}
\frac{\langle \Psi_G\mid\mathcal{H}\mid\Psi_G \rangle}{\langle\Psi_G\mid\Psi_G\rangle} = \langle \Psi_0 \mid \mathcal{\hat H}_{\itext{eff}} \mid \Psi_0 \rangle \equiv \langle \mathcal{\hat H}_{\itext{eff}} \rangle_0,
\end{equation}
where
\begin{multline}
 \mathcal{\hat H}_{\mathit{eff}} = t \sum_{\langle i, j \rangle, \sigma} g_{i\sigma} g_{j\sigma} \left( \hat c^\dagger_{i\sigma} \hat c_{j \sigma} + \mbox{H.c.} \right) \\
+ t'\!\!\! \sum_{\langle\langle i, j \rangle\rangle, \sigma}\!\! g_{i\sigma} g_{j\sigma} \left( \hat c^\dagger_{i\sigma} \hat c_{j \sigma}  + \mbox{H.c.} \right) \\
+  J \sum_{\langle i,\, j\rangle,\, \sigma} g^s_i g^s_j\,\bS_i \cdot \bS_{j} + U \sum_i \hat n_{i\uparrow} \hat n_{i\downarrow}, \label{eq:H_eff_tJU--B}
\end{multline}
where $\sum_{\langle i, j \rangle}$ and $\sum_{\langle\langle i, j \rangle\rangle}$ denotes summation over all unique pairs of first and second nearest neighbors,
$\mbox{H.c.}$ is the Hermitian conjugation,
and $g_{i\sigma}$, $g^s_{i}$ are renormalization factors \cite{Ogawa1975-ProgTheorPhys.53.614, Zhang1988-SuperSciTech.1.36}
 \begin{eqnarray}
 g_{i\sigma} & = & \sqrt{g^s_i} \left( \sqrt{\frac{(1-n_{i\bar\sigma})(1-n+d^2)}{1-n_{i\sigma}}} + \sqrt{\frac{n_{i\bar\sigma}\, d^2}{n_{i\sigma}}} \right), \label{eq:tJU:gt_iSigma--B}\\
 g^{s}_{i} & = & \frac{n-2d^2}{n - 2n_{i\sigma}n_{i\bar\sigma}},  \label{eq:tJU:gs_i--B}
\end{eqnarray}
with $n \equiv \langle \hat n_{i\uparrow} + \hat n_{i\downarrow} \rangle_0$, $d^2 \equiv \langle \hat n_{i\uparrow} \hat n_{i\downarrow} \rangle_0$, and
\begin{equation}
 n_{i\sigma} \equiv \langle \hat c_{i\sigma}^\dagger \hat c_{i\sigma} \rangle_0 \equiv \frac{1}{2} \left( n + \sigma e^{i {\bf Q} \cdot {\bf R}_i}\, m \right), \label{eq:rozpisane_n--B}
\end{equation}
where $m$ is (bare) sublattice magnetization per site, ${\bf Q} \equiv (\pi,\, \pi)$, and  ${\bf R}_i$ is the position vector of site~$i$.
We divide the lattice into two sublattices, $A$ where on average the spin is \emph{up}, and $B$ where on average is \emph{down} (cf.\ Fig.~\ref{sites}).
Thus $n_{i\in A,\sigma} \equiv \frac{1}{2} \left( n + \sigma m \right)$, and $n_{i\in B,\sigma} \equiv \frac{1}{2} \left( n - \sigma m \right)$.

We define average hopping amplitude for the first and the second nearest neighbors (1st and 2nd n.n.) as
\begin{equation}
 \chi_{ij\sigma} \equiv \langle \hat c_{i\sigma}^\dagger \hat c_{j\sigma} \rangle_0 \equiv \left\{
   \begin{array}{ll}
     \chi & \hspace{6pt} \text{for 1st n.n.}, \\
     \chi_S + \sigma e^{i {\bf Q} \cdot \br_i} \chi_T & \hspace{6pt} \text{for 2nd n.n.},
   \end{array}
\right. \label{eq:def-Delta--B}
\end{equation}
where $\chi \equiv \chi_{AB}$ denotes hopping between sublattices $A$ and $B$ (or vice versa, cf.\ left panel in Fig.~\ref{sites}); $\chi_{S} \equiv \frac{1}{2} ( \chi_{AA} + \chi_{BB} )$ and $\chi_{T} \equiv \frac{1}{2} ( \chi_{AA} - \chi_{BB} )$, where $\chi_{AA}$ and $\chi_{BB}$ denotes hopping within one sublattice.
We define also the electron pairing between nearest neighbors as
\begin{equation}
 \Delta_{ij\sigma} \equiv \langle \hat c_{i \sigma} \hat c_{j \bar\sigma} \rangle_0 =
 - \tau_{ij}\left(\sigma \Delta_S + e^{i {\bf Q}\cdot {\bf R}_i} \Delta_T\right),
\label{eq:Delta_box--B}
\end{equation}
where $ \tau_{ij} \equiv 1$ for $j=i\pm \hat x$, and $ \tau_{ij} \equiv -1$ for $j=i\pm \hat y$ to ensure $d$-wave symmetry. $\Delta_S \equiv \frac{1}{4}\left( \Delta_A + \Delta_B + \mbox{H.c.} \right)$ and $\Delta_T \equiv \frac{1}{4} \left( \Delta_A - \Delta_B + \mbox{H.c.} \right)$, cf.\ right panel in Fig.~\ref{sites}.
We assume that all the above averages: $\chi$, $\chi_S$, $\chi_T$, $\Delta_S$, and $\Delta_T$, are real. 
Finally, we are able to calculate the average $W \equiv \langle \mathcal{\hat H} \rangle_0$, which has the form
\begin{multline}
 \frac{W}{\Lambda} = 8 g_t t \chi + 4 g^{max}_{t'} t' \chi_{S} + 4 g^{min}_{t'} t' \chi_{S} \\
+ g_s J \left( -\frac{1}{2} m^2 -3 \chi^2 - 3 \Delta^2_S + \Delta^2_T \right) + U d^2, \label{eq:W}
\end{multline}
where the renormalization factors $g_t \equiv g_{i \in A \sigma} g_{j \in B \sigma}$, $g^{max}_{t'} \equiv g_{i \in A \uparrow} g_{j \in A \uparrow}$, $g^{min}_{t'} \equiv g_{i \in A \downarrow} g_{j \in A \downarrow}$, and $g_s \equiv g^{s}_{i \in A} g^{s}_{j \in B}$.


\section{Statistically-consistent Gutzwiller Approximation}
\label{sec:SGA}

To determine the stable phases and their characteristics (sublattice magnetization, SC gap, etc.) we construct the grand potential functional,
which we next minimize with respect to all parameters. However, to ensure that the averages calculated in a~self-consistent manner are equal
to those obtained variationally, we first use the so-called Statistically-consistent Gutzwiller Approximation (SGA)
(cf.\ introduction to SGA \cite{Jedrak2010-arXiv},
and examples of its use in the context of the $t$--$J$ model \cite{Jedrak2010-prb, *Jedrak2011-PhysRevB.83.104512, Kaczmarczyk2011-prb},
the $t$--$J$--$U$ model \cite{Abram2013-PRB.88.094502}, 
the Anderson–Kondo lattice model \cite{Olga2012-JPhysCM.24.205602, Olga2013-PhysStatusSolidi.250.609},
the extended Hubbard models \cite{Zegrodnik2013-NewJPhys.15.073050, Zegrodnik2013-JPhysCM.25.435601, Kadzielawa2013-EurPhysJB.86.252}, or
the liquid $^3$He \cite{Wysokinski2014-JPCM.26.055601}).
Here, we impose constraints on each average, which is present in Eq.\ (\ref{eq:W}). Hence, our effective Hamiltonian takes the form
\begin{multline}
 \hat K =  W - \sum_{\langle i,j\rangle,\sigma} \left( \lambda_{ij\sigma}^\chi \left( \hat c_{i\sigma}^\dagger \hat c_{j\sigma} - \chi_{ij\sigma} \right) + \text{H.c.} \right) \\
- \sum_{\langle\langle i,j\rangle\rangle,\sigma} \left( \lambda_{ij\sigma}^{\chi} \left( \hat c_{i\sigma}^\dagger \hat c_{j\sigma} - \chi_{ij\sigma} \right) + \text{H.c.} \right) \\
- \sum_{\langle i,j\rangle} \left( \lambda_{ij\sigma}^\Delta \left( \hat c_{i\sigma} \hat c_{j\bar\sigma} - \Delta_{ij\sigma} \right) + \text{H.c.} \right) \\
- \sum_{i\sigma} \left( \lambda_{i\sigma}^n \left( \hat n_{i\sigma} - n_{i\sigma} \right) \right)
- \mu \sum_{i\sigma} \hat n_{i\sigma}, \label{eq:K--B}
\end{multline}
where we have also introduced the chemical potential term $-\mu \sum_{i\sigma} \hat n_{i\sigma}$.
Symbols $\{ \lambda_i \}$ stand for Lagrange multipliers, having the same form as the corresponding to them averages, namely
\begin{subequations}
 \begin{eqnarray}
 \lambda^n_{i\sigma} & = & \frac{1}{2} \left( \lambda_n + \sigma  e^{i {\bf Q} \cdot {\bf R}_i} \lambda_m \right) , \label{eq:lambda1} \\
\lambda_{ij\sigma}^\chi & \equiv & \left\{
   \begin{array}{ll}
     \lambda_\chi & \hspace{6pt} \text{for 1st n.n.}, \\
     \lambda_{\chi_S} + \sigma e^{i {\bf Q} \cdot \br_i} \lambda_{\chi_T} & \hspace{6pt} \text{for 2nd n.n.},
   \end{array}
\right. \label{eq:lambda2} \\
  \lambda^{\Delta}_{ij\sigma} & = & - \tau_{ij} \left( \sigma \lambda_{\Delta_S} + i e^{i {\bf Q} \cdot {\bf R}_i} \lambda_{\Delta_T} \right). \label{eq:lambda3}
 \end{eqnarray}
\end{subequations}

\noindent In the next step we diagonalize the grand Hamiltonian $\hat K$ and construct the grand potential functional ${\mathcal{F} = -\frac{1}{\beta} \ln{\mathcal{Z}}}$,
where $\beta = 1/k_B T$, and $\mathcal{Z} = \mathrm{Tr} \big( e^{-\beta \hat K} \big)$.
The minimization conditions for determining all quantities and Lagrange multiplies are
\begin{equation}
\frac{\partial \mathcal{F}}{\partial A_i} = 0, \hspace{12pt}
\frac{\partial \mathcal{F}}{\partial \lambda_i} = 0, \hspace{12pt}
\frac{\partial \mathcal{F}}{\partial d} = 0.
\end{equation}
where $\{A_i\}$ denote here all $7$ averages: $\chi$, $\chi_S$, $\chi_T$, $\Delta_S$, $\Delta_T$,
$n$, and $m$, while $\{\lambda_i\}$ denote all Lagrange multipliers $\lambda_\chi$, $\lambda_{\chi_S}$, $\lambda_{\chi_T}$, $\lambda_{\Delta_S}$, $\lambda_{\Delta_T}$,
$\lambda_n$, and $\lambda_m$.
The system of equations is solved self-consistently.
To determine the stability of physical phases, free energy has to be calculated according to the prescription
\begin{equation}
 F =  \mathcal{F}_0 + \Lambda \mu n, \label{eq:praw_energia_GA--B}
\end{equation}
where $\mathcal{F}_0$ is the value of the grand potential functional $\mathcal{F}$ at minimum, and $\Lambda$ is the number of lattice sites.


\section{Results}
\label{sec:Results}

The numerical calculations were carried out using GNU Scientific Library (GSL) \cite{GSL-manual} for a~two dimensional, square lattice of $\Lambda = 512 \times 512$ size, and unless stated otherwise, $t=-1$, $J=|t|/3$, and $\beta|t| = 1500$ (it was checked that for such large $\beta \equiv 1/{k_B T}$ we have effectively $T = 0$).

Here, $\chi$, $\chi_S$, $\chi_T$, $\Delta_S$, $\Delta_T$, and $m$ are bare averages. Renormalized by a~proper Gutzwiller factors, they become order parameters of the corresponding phases. Thus: $\chi^c \equiv g_t \chi$, $\chi_S^c \equiv g_{t'} \chi_S$, $\chi_T^c \equiv g_{t'} \chi_T$, $\Delta_S^c \equiv g_\Delta \Delta_S$, $\Delta_T^c \equiv g_\Delta \Delta_T$, and $m^c = g_m m$, where (cf.~Eqs.~\bref{eq:tJU:gt_iSigma--B} and~\bref{eq:tJU:gs_i--B}),
$g_{t} \equiv g_{i \in A \sigma} g_{j \in B \sigma}$,
$g_{t'} \equiv \frac{1}{2} ( g_{i \in A \uparrow} g_{j \in A \uparrow} + g_{i \in A \downarrow} g_{j \in A \downarrow})$,
$g_\Delta \equiv \frac{1}{2} \left( g_{i \in A \uparrow} g_{i \in B \downarrow} + g_{i \in A \downarrow} g_{i \in B \uparrow} \right)$, and
$g_m \equiv g^{s}_{i \in A} g^{s}_{j \in B}$.

\begin{figure}
 \centering
  \includegraphics[width=1 \columnwidth]{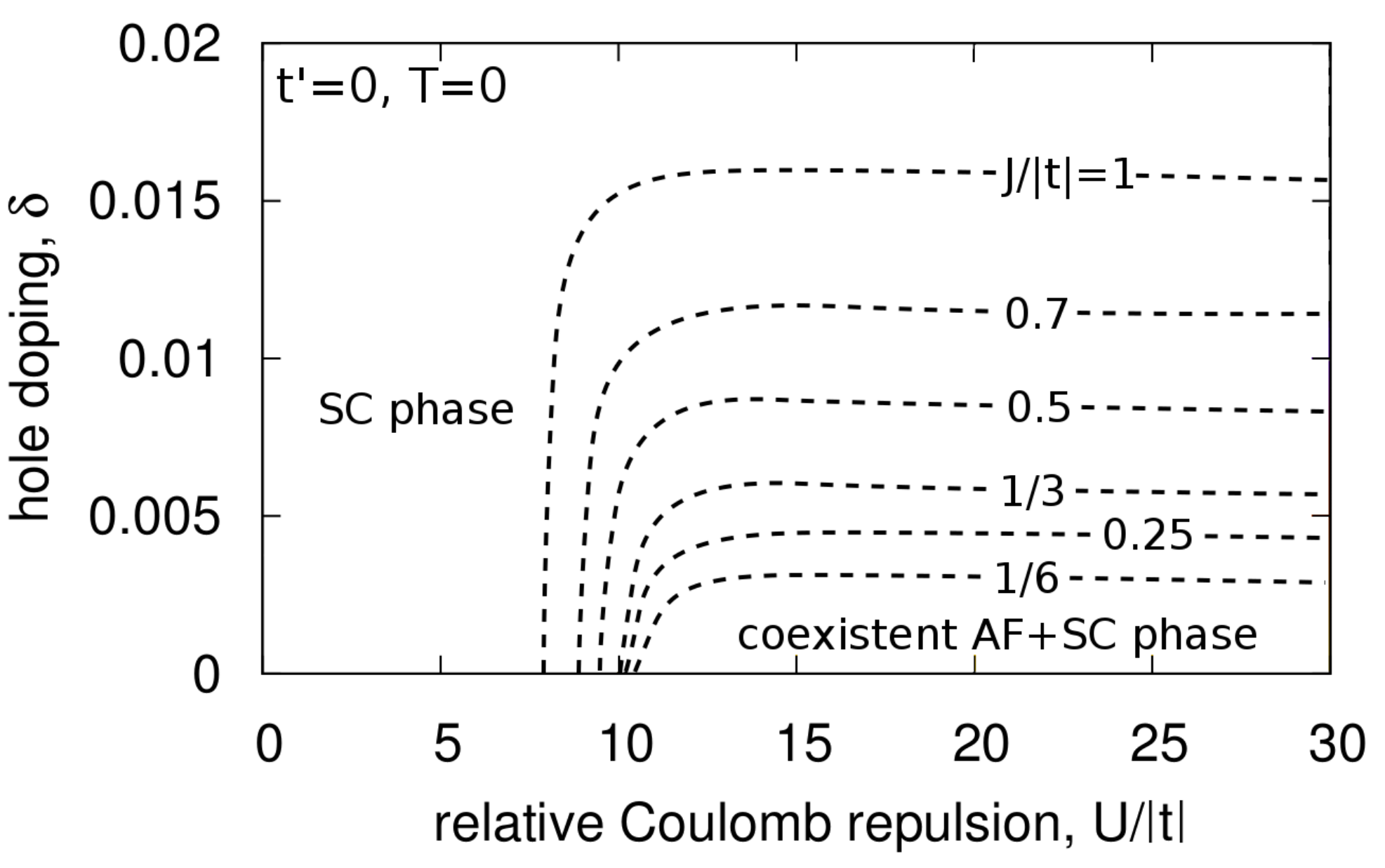}
  \caption{The AF+SC coexistence region for $t'=0$, $T = 0$, and different values of the the exchange coupling $J$ (in units of $t$).} \label{fig:AF+SC_differentJ}
  \end{figure}
  
  \begin{figure}
  \includegraphics[width=0.48\columnwidth]{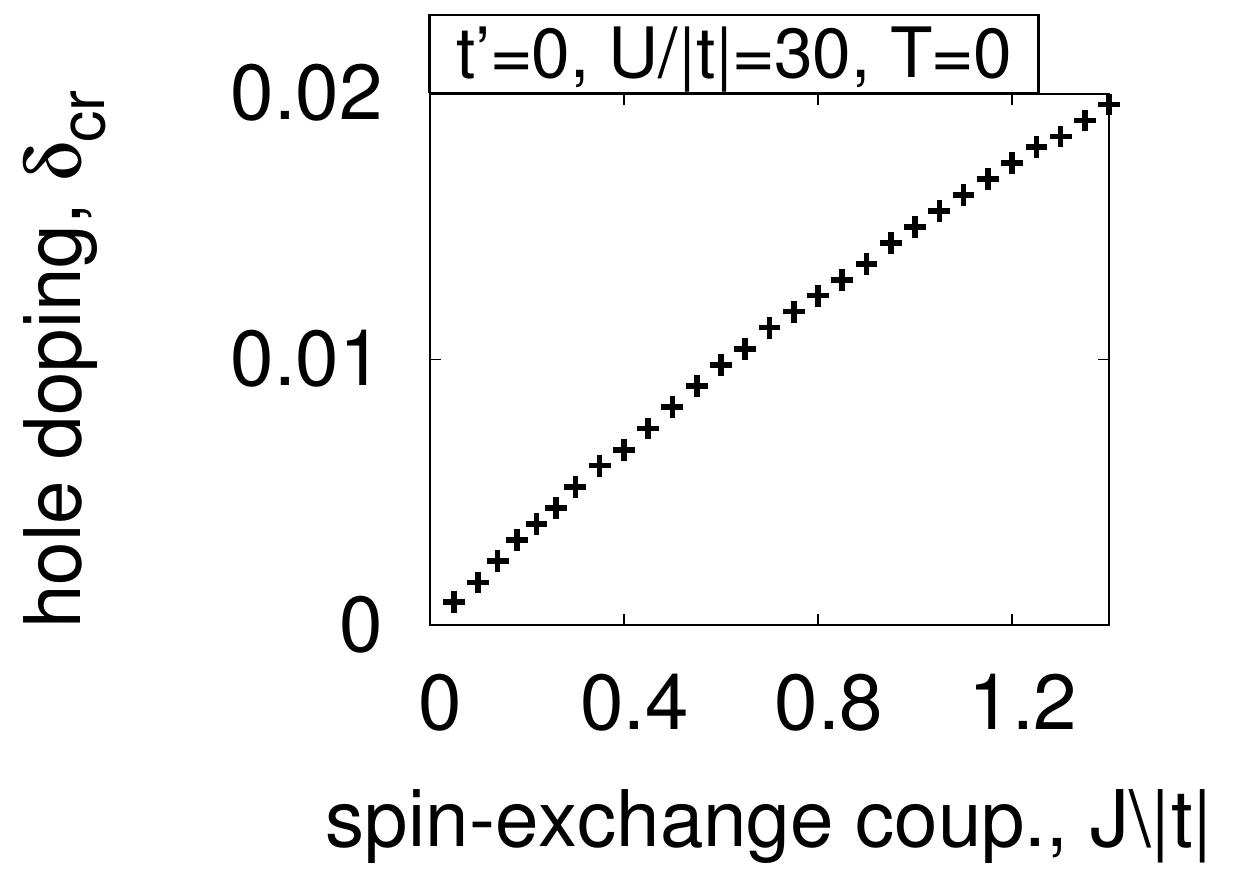} 
  \includegraphics[width=0.48\columnwidth]{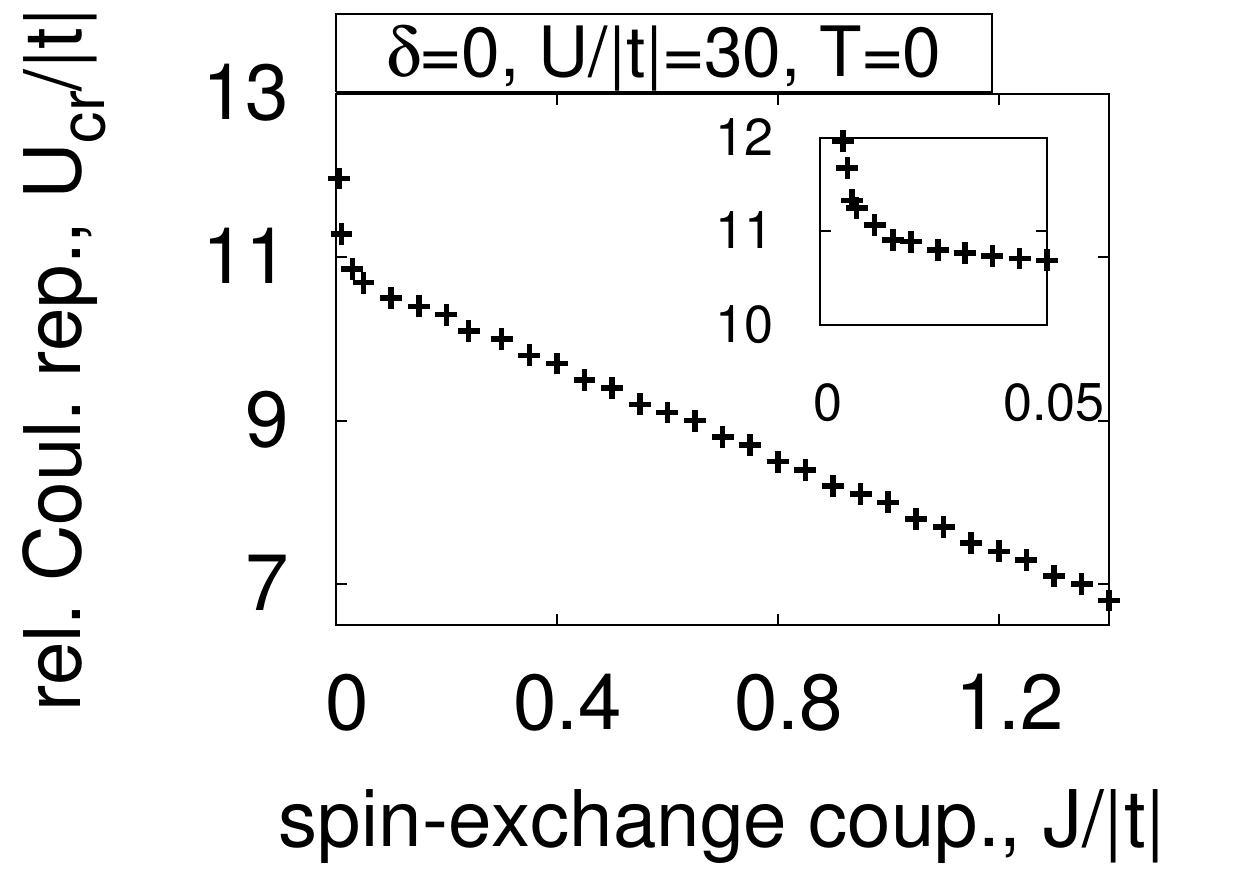}
 \caption{In the \underline{left} panel, the effect of the spin-exchange coupling $J$ on the critical hole doping ($\delta_{cr}$). In the \underline{right} panel, the effect of $J$ on the critical relative Coulomb repulsion ($U_{cr}$). Note, that $\delta_{cr}(J)$ is quasi-linear in the whole range of the tested parameter, while for $U_{cr}(J)$ we observe non-linear behavior for $J/|t|< 0.03$ (cf. the inset in the right panel).}\label{fig:AF+SC_differentJ-panels}
\end{figure}

 \begin{figure}
  \centering
   \includegraphics[width=1 \columnwidth]{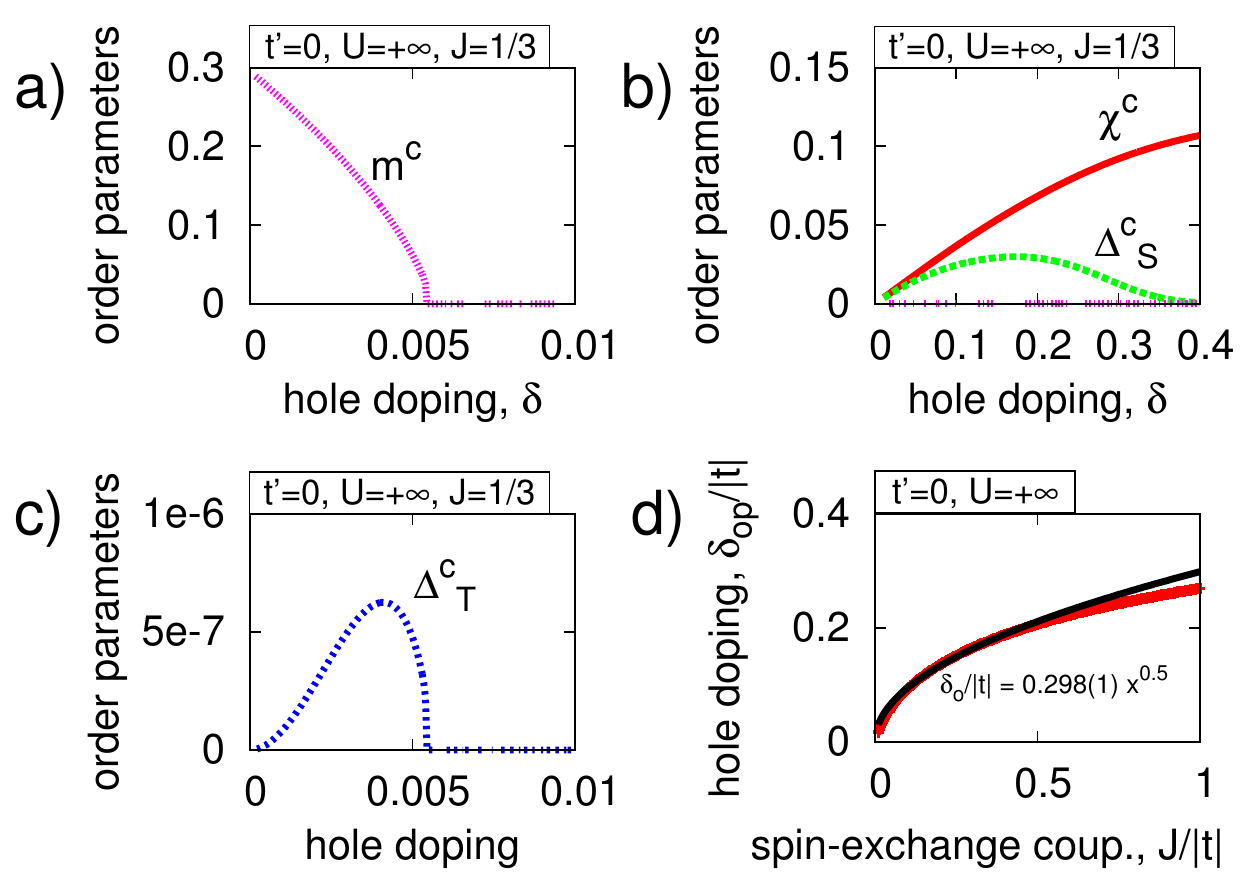}
  \caption{(Color online) In the panels $a)$, $b)$, $c)$, selected order parameters as a~function of doping $\delta$ are presented. Note, that $\Delta_T \neq 0$ only if $m^c \neq 0$. In the panel $d)$, the optimal doping for a~singled SC gap ($\Delta^c_S$) is shown, as a~function of the exchange coupling $J$, in $U \rightarrow + \infty$ limit (red line). The black line is a~numerical fit, $f(x) = 0.298(1)\, x^{0.5}$.}\label{pic:delta-optimal}\label{pic:u-100-panel}
 \end{figure}


\subsection{Results for $t$--$J$--$U$ model, for $t'=0$}

In the limit of the low temperature ($T \rightarrow 0$, i.e.\ $\beta \rightarrow +\infty$) the SC phase is stable for any value of ${\delta>0}$, $U>0$, or $J>0$. For sufficiently large Coulomb repulsion ($U>U_{cr}$) and for small hole doping ($\delta<\delta_{cr}$), a~coexistent AF+SC phase can be found (cf.\ Fig.~\ref{fig:AF+SC_differentJ}).  
For $\delta = 0$ and for $U>U_{cr}$ we obtain the Mott insulating state. For $\delta = 0$ and $U<U_{cr}$ electrons can have double occupancies ($d^2 \neq 0$) and the superconducting pairing is maintained (such a~feature in literature is called the gossamer superconductivity \cite{Laughlin2006}).

The influence of the spin-exchange coupling $J$ on the range of the coexistence region AF+SC was examined. 
$\delta_{cr}$ is a~linear function of of $J$ (cf.\ the left panel in Fig.~\ref{fig:AF+SC_differentJ-panels}), while the critical Coulomb repulsion $U_{cr}$ has non-linear behavior for $J/|t|<0.03$ (the value of $U_{cr}$ grows rapidly when $J$ decrease, cf.\ the right panel in Fig.~\ref{fig:AF+SC_differentJ-panels}).


For $U \rightarrow +\infty$ we reproduce the results of the $t$--$J$ model. As was checked, even for not too large $U$ the convergence to $t$--$J$ model results is sufficient.
For instance, for $U=30$ our results match those for the $t$--$J$ model (so the limit $U = +\infty$) within less than $1\%$ error, and for $U=100$ within an error of less than $0.1\%$.
In Figure~\ref{pic:u-100-panel} in panels $a)$, $b)$, $c)$, the correlated states $\chi^c$, $\Delta_S^c$, $\Delta_T^c$, $m^c$, and $d^2$ are presented for $U=100$ and $\beta|t| = 1500$
(effectively $U = +\infty$ and $T = 0$). Note, that the staggered component of the superconducting gap
($\Delta_T)$ is very small and appears only when $m^c \neq 0$,
i.e., in the AF+SC phase.
However, $\Delta_T$ value is very small when compared to value of $\Delta_S$ (there is $\Delta_T^c/\Delta_S^c < 10^{-4}$),
thus its effect can be practically neglected \footnote{The free energy $F_0$ in minimum (for $T=0$) is equal to $W$ (cf. Eq.~\ref{eq:W}). If $\Delta_T^c/\Delta_S^c \equiv \Delta_T/\Delta_S < 10^{-4}$ then the impact of $\Delta^c_T$ for the final energy of the solution is about $10^{-8}$ smaller than the impact of $\Delta^c_S$. Thus $\Delta_T$ in practical calculations can be neglected.}.

In the last panel $d)$ in Fig.\ \ref{pic:delta-optimal} we show (red line) the optimal doping $\delta_{op}$ for singled SC gap ($\Delta^c_S$) as a~function of $J$. The black line in the panel is a~function $f \sim \sqrt{J/|t|}$, numerically fitted to the data.


 \begin{figure}
  \centering
 \includegraphics[width=1 \columnwidth]{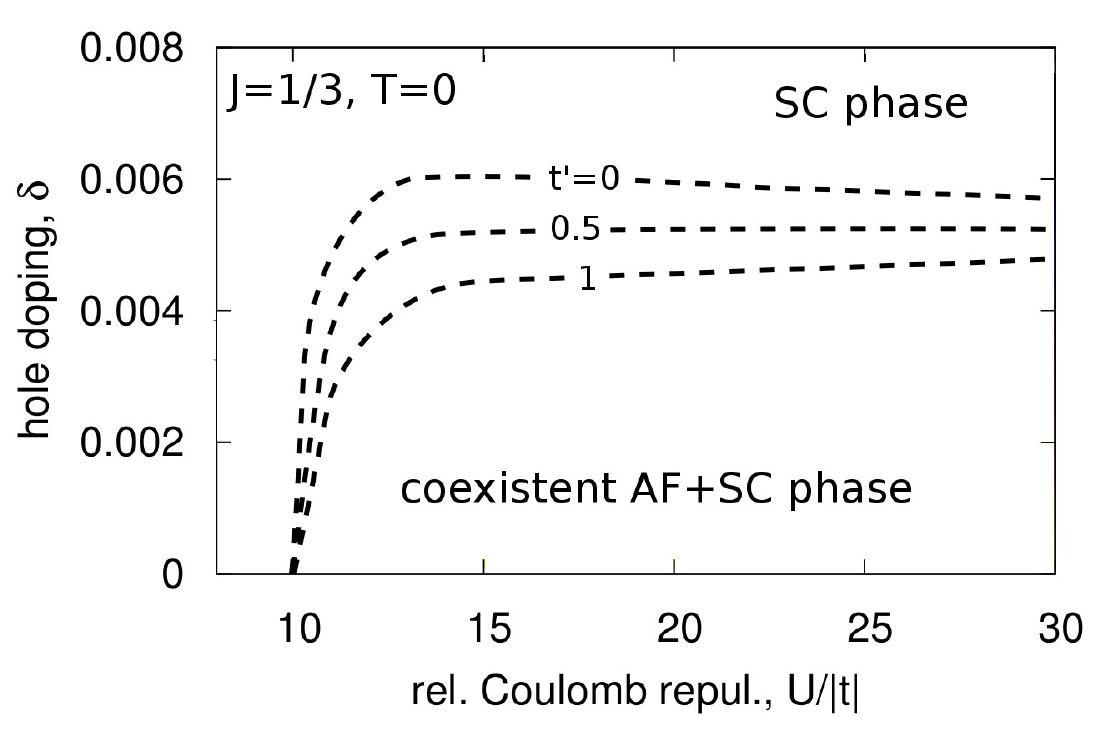}
   \caption{Significance of the second nearest neighbors hopping. Values of $t'$ are given in units of $t$. The presence of $t'$ does not change the AF+SC range in qualitative manner.} \label{pic:2D-tPrim}\vspace{12pt}\label{pic:small-tPrim}
 \end{figure}
 
%

\subsection{A significance of the second nearest neighbors hopping $t'$}

The influence of the second nearest neighbors hopping term $t'$ is exhibited in Fig.~\ref{pic:2D-tPrim}. Note, that the critical Coulomb repulsion for AF+SC phase ($U_{cr}$) is practically independent on the value of $t'$ (it was checked, $U_{cr}(t'=0)$ and $U_{cr}(t'=1)$ differ about $1\%$).
The critical doping ($\delta_{cr}$) is more susceptible to the value of $t'$, but note that the typical value of the $t'$ ranges from $-0.1t$ to $-0.5t$ (cf. Ref.~\onlinecite{Plakida}, Chap. 7.1.2), and in such a~range $\delta_{cr}$ changes only about $10\%$.

 
  \begin{figure}
  \centering
  \includegraphics[width=1 \columnwidth]{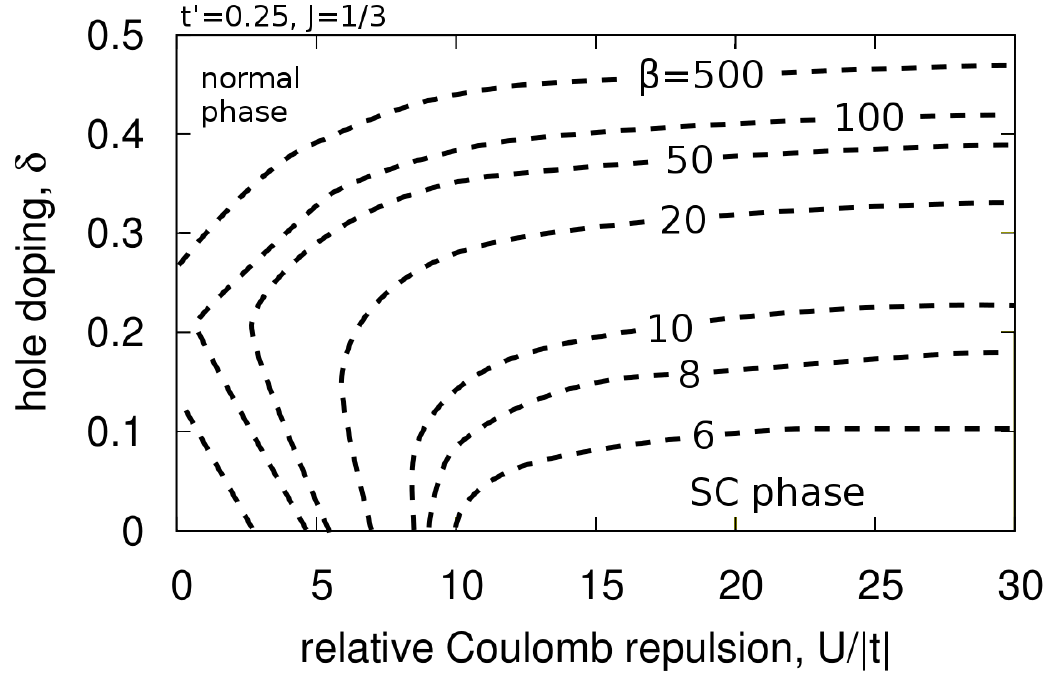}
  \caption{The effect of the temperature (meassured in units of $|t|$) on the stability of SC phase in $t$--$t'$--$J$--$U$ model ($t=-1$, $t'=0.25$). The dashed lines correspond to the range of SC phase for $\beta=500$ (${T \sim 5}$--$12$~K), $\beta=100$ ($25$--$60$~K), $\beta=50$ ($50$--$120$~K), $\beta=20$ ($130$--$290$~K), $\beta=10$ ($250$--$580$~K), $\beta=8$ ($320$--$720$~K), $\beta=6$ ($420$--$1000$~K).} \label{pic:2D-Bdependence}
 \end{figure}
 
\subsection{Non-zero temperatures}

In the limit of the zero temperature, for small $U$ or/and large $\delta$, the value of the SC order parameter $\Delta^c_S$ is small, but still nonzero. Increasing the temperature (decreasing the parameter $\beta$), the paramagnetic (PM) phase appears in region where the order parameter of SC phase was weak (cf. Fig.\ \ref{pic:2D-Bdependence}). For large $T$ (small $\beta$), the range of the SC phase is reduced to the vicinity of the Mott-insulator phase ($\delta \gtrsim 0$, and $U>U_{cr}$).

The measured value of the hopping term $t$ for the cuprates ranges from $0.22$~eV to $0.5$~eV (cf. Ref.~\onlinecite{Plakida}, Chap. 7.1.2).
Hence the $\beta|t| = 1500$ corresponds to the temperature $2\mbox{--}4$~K,
$\beta|t| = 500$ to $5\mbox{--}12$~K, 
$\beta|t| = 100$ to $25\mbox{--}60$~K,
$\beta|t| = 50$  to $50\mbox{--}120$~K,
$\beta|t| = 20$  to $130\mbox{--}290$~K,
$\beta|t| = 10$  to $250\mbox{--}580$~K,
$\beta|t| = 8$   to $320\mbox{--}720$~K,
$\beta|t| = 6$   to $420\mbox{--}1000$~K.

 
\section{Conclusions}
\label{sec:Conclusions}

In this work, the $t$--$t'$--$J$--$U$ model was studied in the SGA scheme which plays the role of the mean-field approximation.
In the limit of the zero temperature, three phases were found: superconductivity (SC), coexistent antiferromagnetic-superconducting state (AF+SC),
and the Mott-insulating phase (for the half filling). The AF+SC phase exists only for sufficiently large Coulomb repulsion ($U>U_{cr}$) and
for small hole doping ($\delta<\delta_{cr}$). We have shown how the range of AF+SC coexistence varies with $J$ and $t'$. The impact of $J$ was significant,
both for $U_{cr}$ and for $\delta_{cr}$. However, the impact of $t'$ was much smaller and in the range of physical values (for cuprates $t' \sim 0.1\mbox{--}0.5 |t|$), it can be marginal.

The impact of the non-zero temperatures was tested. For $T>0$, additionally to SC and AF+SC phases, a~paramagnetic phase (normal phase) appears.
The ranges of SC and AF+SC phases decrease with the temperature, but they remain stable even for relatively high temperature (${\sim{}\!\!1000}$~K).
Such results, contradictory to the experiments, can be explained by the used method (the saddle-point method) and approximations used (the mean-field and the Gutzwiller approximation).
To study more accurately the stability of the phases, more sophisticated method should be used (cf. e.g. the diagrammatic expansion for Gutzwiller-wave functions (DE-GWF) \cite{Kaczmarczyk2013-PRB.88.115127}).


\section*{Acknowledgments}
\label{sec:Acknowledgments}

I would like to express my gratitude to Prof. J\'ozef Spa\l{}ek for his support and helpful detailed comments.
I would also like to thank  Marcin Wysoki\'nski for discussions, and Allison Hartnett for her critical reading of the manuscript.
This research was supported by the Foundation for Polish Science (FNP) under the Grant TEAM.
Parts of the calculations were performed on the TERAACMIN supercomputer in the Academic Centre for Materials and Na\-no\-tech\-no\-lo\-gy (ACMIN)
of AGH University of Science and Technology in Krak\'ow.

\bibliographystyle{apsrev}
\bibliography{tJU-ActaPolonicaA.bib}

\begin{thebibliography}{33}
\expandafter\ifx\csname natexlab\endcsname\relax\def\natexlab#1{#1}\fi
\expandafter\ifx\csname bibnamefont\endcsname\relax
  \def\bibnamefont#1{#1}\fi
\expandafter\ifx\csname bibfnamefont\endcsname\relax
  \def\bibfnamefont#1{#1}\fi
\expandafter\ifx\csname citenamefont\endcsname\relax
  \def\citenamefont#1{#1}\fi
\expandafter\ifx\csname url\endcsname\relax
  \def\url#1{\texttt{#1}}\fi
\expandafter\ifx\csname urlprefix\endcsname\relax\def\urlprefix{URL }\fi
\providecommand{\bibinfo}[2]{#2}
\providecommand{\eprint}[2][]{\url{#2}}

\bibitem[{\citenamefont{Spałek and
  Oleś}(1977)}]{Spatek1977-PhysicaBC.86-88.375}
\bibinfo{author}{\bibfnamefont{J.}~\bibnamefont{Spałek}} \bibnamefont{and}
  \bibinfo{author}{\bibfnamefont{A.}~\bibnamefont{Oleś}},
  \bibinfo{journal}{Physica B+C} \textbf{\bibinfo{volume}{86–88}},
  \bibinfo{pages}{375 } (\bibinfo{year}{1977}).

\bibitem[{\citenamefont{Chao et~al.}(1977)\citenamefont{Chao, Spałek, and
  Oleś}}]{ChaoSpalekOles1977-JPhysC.10.L271}
\bibinfo{author}{\bibfnamefont{K.~A.} \bibnamefont{Chao}},
  \bibinfo{author}{\bibfnamefont{J.}~\bibnamefont{Spałek}}, \bibnamefont{and}
  \bibinfo{author}{\bibfnamefont{A.~M.} \bibnamefont{Oleś}},
  \bibinfo{journal}{J. Phys. C} \textbf{\bibinfo{volume}{10}},
  \bibinfo{pages}{L271} (\bibinfo{year}{1977}).

\bibitem[{\citenamefont{Spałek}(2007)}]{Spalek2007-ActaPhysPolonA.111.409}
\bibinfo{author}{\bibfnamefont{J.}~\bibnamefont{Spałek}},
  \bibinfo{journal}{Acta Phys. Polon. A} \textbf{\bibinfo{volume}{111}},
  \bibinfo{pages}{409} (\bibinfo{year}{2007}).

\bibitem[{\citenamefont{Dagotto}(1994)}]{Dagotto1994-RevModPhys.66.763}
\bibinfo{author}{\bibfnamefont{E.}~\bibnamefont{Dagotto}},
  \bibinfo{journal}{Rev. Mod. Phys.} \textbf{\bibinfo{volume}{66}},
  \bibinfo{pages}{763} (\bibinfo{year}{1994}).

\bibitem[{\citenamefont{Lee et~al.}(2006)\citenamefont{Lee, Nagaosa, and
  Wen}}]{Lee2006-RevModPhys.78.17}
\bibinfo{author}{\bibfnamefont{P.~A.} \bibnamefont{Lee}},
  \bibinfo{author}{\bibfnamefont{N.}~\bibnamefont{Nagaosa}}, \bibnamefont{and}
  \bibinfo{author}{\bibfnamefont{X.-G.} \bibnamefont{Wen}},
  \bibinfo{journal}{Rev. Mod. Phys.} \textbf{\bibinfo{volume}{78}},
  \bibinfo{pages}{17} (\bibinfo{year}{2006}).

\bibitem[{\citenamefont{Jędrak}(2011)}]{Jedrak2011-PhD}
\bibinfo{author}{\bibfnamefont{J.}~\bibnamefont{Jędrak}}, Ph.D. thesis,
  \bibinfo{school}{Jagiellonian University, Kraków} (\bibinfo{year}{2011}),
  \urlprefix\url{http://th-www.if.uj.edu.pl/ztms/download/phdTheses/Jakub_Jedrak_doktorat.pdf}.

\bibitem[{\citenamefont{Lin}(1991)}]{Lin1991-PRB.44.4674}
\bibinfo{author}{\bibfnamefont{H.~Q.} \bibnamefont{Lin}},
  \bibinfo{journal}{Phys. Rev. B} \textbf{\bibinfo{volume}{44}},
  \bibinfo{pages}{4674} (\bibinfo{year}{1991}).

\bibitem[{\citenamefont{Zhang and Rice}(1988)}]{Zhang1988-PhysRevB.37.3759}
\bibinfo{author}{\bibfnamefont{F.~C.} \bibnamefont{Zhang}} \bibnamefont{and}
  \bibinfo{author}{\bibfnamefont{T.~M.} \bibnamefont{Rice}},
  \bibinfo{journal}{Phys. Rev. B} \textbf{\bibinfo{volume}{37}},
  \bibinfo{pages}{3759} (\bibinfo{year}{1988}).

\bibitem[{\citenamefont{Zhang}(2003)}]{Zhang2003-prl}
\bibinfo{author}{\bibfnamefont{F.~C.} \bibnamefont{Zhang}},
  \bibinfo{journal}{Phys. Rev. Lett.} \textbf{\bibinfo{volume}{90}},
  \bibinfo{pages}{207002} (\bibinfo{year}{2003}).

\bibitem[{\citenamefont{Gan et~al.}(2005{\natexlab{a}})\citenamefont{Gan,
  Zhang, and Su}}]{Gan2005-prb}
\bibinfo{author}{\bibfnamefont{J.~Y.} \bibnamefont{Gan}},
  \bibinfo{author}{\bibfnamefont{F.~C.} \bibnamefont{Zhang}}, \bibnamefont{and}
  \bibinfo{author}{\bibfnamefont{Z.~B.} \bibnamefont{Su}},
  \bibinfo{journal}{Phys. Rev. B} \textbf{\bibinfo{volume}{71}},
  \bibinfo{pages}{014508} (\bibinfo{year}{2005}{\natexlab{a}}).

\bibitem[{\citenamefont{Gan et~al.}(2005{\natexlab{b}})\citenamefont{Gan, Chen,
  Su, and Zhang}}]{Gan2005-prl}
\bibinfo{author}{\bibfnamefont{J.~Y.} \bibnamefont{Gan}},
  \bibinfo{author}{\bibfnamefont{Y.}~\bibnamefont{Chen}},
  \bibinfo{author}{\bibfnamefont{Z.~B.} \bibnamefont{Su}}, \bibnamefont{and}
  \bibinfo{author}{\bibfnamefont{F.~C.} \bibnamefont{Zhang}},
  \bibinfo{journal}{Phys. Rev. Lett.} \textbf{\bibinfo{volume}{94}},
  \bibinfo{pages}{067005} (\bibinfo{year}{2005}{\natexlab{b}}).

\bibitem[{\citenamefont{Bernevig et~al.}(2003)\citenamefont{Bernevig, Laughlin,
  and Santiago}}]{Bernevig-PhysRevLett.91.147003}
\bibinfo{author}{\bibfnamefont{B.~A.} \bibnamefont{Bernevig}},
  \bibinfo{author}{\bibfnamefont{R.~B.} \bibnamefont{Laughlin}},
  \bibnamefont{and} \bibinfo{author}{\bibfnamefont{D.~I.}
  \bibnamefont{Santiago}}, \bibinfo{journal}{Phys. Rev. Lett.}
  \textbf{\bibinfo{volume}{91}}, \bibinfo{pages}{147003}
  (\bibinfo{year}{2003}).

\bibitem[{Note1()}]{Note1}
Note1, \bibinfo{note}{some attempts was made by some authors, cf.\ PRB
  {\protect \bf 71}, 104505 (2005), PRA {\protect \bf 79} 063611 (2009), and J.
  Phys.: Condens Matter {\protect \bf 23}, 495602 (2011), but their methods
  suffered from some serious inconsistencies which affected their final results
  (cf.\ discussion in Ref.~\protect \onlinecite {Abram2013-PRB.88.094502}).}

\bibitem[{\citenamefont{Abram et~al.}(2013)\citenamefont{Abram, Kaczmarczyk,
  J\k{e}drak, and Spa\l{}ek}}]{Abram2013-PRB.88.094502}
\bibinfo{author}{\bibfnamefont{M.}~\bibnamefont{Abram}},
  \bibinfo{author}{\bibfnamefont{J.}~\bibnamefont{Kaczmarczyk}},
  \bibinfo{author}{\bibfnamefont{J.}~\bibnamefont{J\k{e}drak}},
  \bibnamefont{and}
  \bibinfo{author}{\bibfnamefont{J.}~\bibnamefont{Spa\l{}ek}},
  \bibinfo{journal}{Phys. Rev. B} \textbf{\bibinfo{volume}{88}},
  \bibinfo{pages}{094502} (\bibinfo{year}{2013}).

\bibitem[{\citenamefont{Gutzwiller}(1963)}]{Gutzwiller1963-prl}
\bibinfo{author}{\bibfnamefont{M.~C.} \bibnamefont{Gutzwiller}},
  \bibinfo{journal}{Phys. Rev. Lett.} \textbf{\bibinfo{volume}{10}},
  \bibinfo{pages}{159} (\bibinfo{year}{1963}).

\bibitem[{\citenamefont{Gutzwiller}(1965)}]{Gutzwiller1965-pr}
\bibinfo{author}{\bibfnamefont{M.~C.} \bibnamefont{Gutzwiller}},
  \bibinfo{journal}{Phys. Rev.} \textbf{\bibinfo{volume}{137}},
  \bibinfo{pages}{A1726} (\bibinfo{year}{1965}).

\bibitem[{\citenamefont{Ogawa et~al.}(1975)\citenamefont{Ogawa, Kanda, and
  Matsubara}}]{Ogawa1975-ProgTheorPhys.53.614}
\bibinfo{author}{\bibfnamefont{T.}~\bibnamefont{Ogawa}},
  \bibinfo{author}{\bibfnamefont{K.}~\bibnamefont{Kanda}}, \bibnamefont{and}
  \bibinfo{author}{\bibfnamefont{T.}~\bibnamefont{Matsubara}},
  \bibinfo{journal}{Prog. Theor. Phys.} \textbf{\bibinfo{volume}{53}},
  \bibinfo{pages}{614} (\bibinfo{year}{1975}).

\bibitem[{\citenamefont{Zhang et~al.}(1988)\citenamefont{Zhang, Gros, Rice, and
  Shiba}}]{Zhang1988-SuperSciTech.1.36}
\bibinfo{author}{\bibfnamefont{F.~C.} \bibnamefont{Zhang}},
  \bibinfo{author}{\bibfnamefont{C.}~\bibnamefont{Gros}},
  \bibinfo{author}{\bibfnamefont{T.~M.} \bibnamefont{Rice}}, \bibnamefont{and}
  \bibinfo{author}{\bibfnamefont{H.}~\bibnamefont{Shiba}},
  \bibinfo{journal}{Supercond. Sci. Technol.} \textbf{\bibinfo{volume}{1}},
  \bibinfo{pages}{36} (\bibinfo{year}{1988}).

\bibitem[{\citenamefont{J\k{e}drak et~al.}(2010)\citenamefont{J\k{e}drak,
  Kaczmarczyk, and Spa\l{}ek}}]{Jedrak2010-arXiv}
\bibinfo{author}{\bibfnamefont{J.}~\bibnamefont{J\k{e}drak}},
  \bibinfo{author}{\bibfnamefont{J.}~\bibnamefont{Kaczmarczyk}},
  \bibnamefont{and}
  \bibinfo{author}{\bibfnamefont{J.}~\bibnamefont{Spa\l{}ek}},
  \bibinfo{journal}{arXiv:cond-mat/1008.0021}  (\bibinfo{year}{2010}),
  \bibinfo{note}{unpublished}.

\bibitem[{\citenamefont{Jędrak and Spa\l{}ek}(2010)}]{Jedrak2010-prb}
\bibinfo{author}{\bibfnamefont{J.}~\bibnamefont{Jędrak}} \bibnamefont{and}
  \bibinfo{author}{\bibfnamefont{J.}~\bibnamefont{Spa\l{}ek}},
  \bibinfo{journal}{Phys. Rev. B} \textbf{\bibinfo{volume}{81}},
  \bibinfo{pages}{073108} (\bibinfo{year}{2010}).

\bibitem[{\citenamefont{J\ifmmode~\mbox{\k{e}}\else \k{e}\fi{}drak and
  Spa\l{}ek}(2011)}]{Jedrak2011-PhysRevB.83.104512}
\bibinfo{author}{\bibfnamefont{J.}~\bibnamefont{J\ifmmode~\mbox{\k{e}}\else
  \k{e}\fi{}drak}} \bibnamefont{and}
  \bibinfo{author}{\bibfnamefont{J.}~\bibnamefont{Spa\l{}ek}},
  \bibinfo{journal}{Phys. Rev. B} \textbf{\bibinfo{volume}{83}},
  \bibinfo{pages}{104512} (\bibinfo{year}{2011}).

\bibitem[{\citenamefont{Kaczmarczyk and Spa\l{}ek}(2011)}]{Kaczmarczyk2011-prb}
\bibinfo{author}{\bibfnamefont{J.}~\bibnamefont{Kaczmarczyk}} \bibnamefont{and}
  \bibinfo{author}{\bibfnamefont{J.}~\bibnamefont{Spa\l{}ek}},
  \bibinfo{journal}{Phys. Rev. B} \textbf{\bibinfo{volume}{84}},
  \bibinfo{pages}{125140} (\bibinfo{year}{2011}).

\bibitem[{\citenamefont{Howczak and
  Spa{\l}ek}(2012)}]{Olga2012-JPhysCM.24.205602}
\bibinfo{author}{\bibfnamefont{O.}~\bibnamefont{Howczak}} \bibnamefont{and}
  \bibinfo{author}{\bibfnamefont{J.}~\bibnamefont{Spa{\l}ek}},
  \bibinfo{journal}{J. Phys.: Condens. Matter} \textbf{\bibinfo{volume}{24}},
  \bibinfo{pages}{205602} (\bibinfo{year}{2012}).

\bibitem[{\citenamefont{Howczak et~al.}(2013)\citenamefont{Howczak,
  Kaczmarczyk, and Spa{\l}ek}}]{Olga2013-PhysStatusSolidi.250.609}
\bibinfo{author}{\bibfnamefont{O.}~\bibnamefont{Howczak}},
  \bibinfo{author}{\bibfnamefont{J.}~\bibnamefont{Kaczmarczyk}},
  \bibnamefont{and}
  \bibinfo{author}{\bibfnamefont{J.}~\bibnamefont{Spa{\l}ek}},
  \bibinfo{journal}{Phys. Status Solidi (b)} \textbf{\bibinfo{volume}{250}},
  \bibinfo{pages}{609} (\bibinfo{year}{2013}), ISSN \bibinfo{issn}{1521-3951}.

\bibitem[{\citenamefont{Zegrodnik et~al.}(2013)\citenamefont{Zegrodnik,
  Spa{\l}ek, and B\"{u}nemann}}]{Zegrodnik2013-NewJPhys.15.073050}
\bibinfo{author}{\bibfnamefont{M.}~\bibnamefont{Zegrodnik}},
  \bibinfo{author}{\bibfnamefont{J.}~\bibnamefont{Spa{\l}ek}},
  \bibnamefont{and}
  \bibinfo{author}{\bibfnamefont{J.}~\bibnamefont{B\"{u}nemann}},
  \bibinfo{journal}{New J. Phys.} \textbf{\bibinfo{volume}{15}},
  \bibinfo{pages}{073050} (\bibinfo{year}{2013}).

\bibitem[{\citenamefont{Spa{\l}ek and
  Zegrodnik}(2013)}]{Zegrodnik2013-JPhysCM.25.435601}
\bibinfo{author}{\bibfnamefont{J.}~\bibnamefont{Spa{\l}ek}} \bibnamefont{and}
  \bibinfo{author}{\bibfnamefont{M.}~\bibnamefont{Zegrodnik}},
  \bibinfo{journal}{J. Phys.: Condens. Matter} \textbf{\bibinfo{volume}{25}},
  \bibinfo{pages}{435601} (\bibinfo{year}{2013}).

\bibitem[{\citenamefont{K\k{a}dzielawa
  et~al.}(2013)\citenamefont{K\k{a}dzielawa, Spa{\l}ek, Kurzyk, and
  W\'{o}jcik}}]{Kadzielawa2013-EurPhysJB.86.252}
\bibinfo{author}{\bibfnamefont{A.~P.} \bibnamefont{K\k{a}dzielawa}},
  \bibinfo{author}{\bibfnamefont{J.}~\bibnamefont{Spa{\l}ek}},
  \bibinfo{author}{\bibfnamefont{J.}~\bibnamefont{Kurzyk}}, \bibnamefont{and}
  \bibinfo{author}{\bibfnamefont{W.}~\bibnamefont{W\'{o}jcik}},
  \bibinfo{journal}{Eur. Phys. J. B} \textbf{\bibinfo{volume}{86}},
  \bibinfo{pages}{252} (\bibinfo{year}{2013}), ISSN \bibinfo{issn}{1434-6028}.

\bibitem[{\citenamefont{Wysoki\'{n}ski and
  Spa{\l}ek}(2014)}]{Wysokinski2014-JPCM.26.055601}
\bibinfo{author}{\bibfnamefont{M.~M.} \bibnamefont{Wysoki\'{n}ski}}
  \bibnamefont{and}
  \bibinfo{author}{\bibfnamefont{J.}~\bibnamefont{Spa{\l}ek}},
  \bibinfo{journal}{J. Phys.: Condens. Matter} \textbf{\bibinfo{volume}{26}},
  \bibinfo{pages}{055601} (\bibinfo{year}{2014}).

\bibitem[{\citenamefont{Galassi et~al.}(2009)\citenamefont{Galassi, Davies,
  Theiler, Gough, Jungman, Booth, and Rossi}}]{GSL-manual}
\bibinfo{author}{\bibfnamefont{M.}~\bibnamefont{Galassi}},
  \bibinfo{author}{\bibfnamefont{J.}~\bibnamefont{Davies}},
  \bibinfo{author}{\bibfnamefont{J.}~\bibnamefont{Theiler}},
  \bibinfo{author}{\bibfnamefont{B.}~\bibnamefont{Gough}},
  \bibinfo{author}{\bibfnamefont{P.}~\bibnamefont{Jungman},
  \bibfnamefont{G.~abd~Alken}},
  \bibinfo{author}{\bibfnamefont{M.}~\bibnamefont{Booth}}, \bibnamefont{and}
  \bibinfo{author}{\bibfnamefont{F.}~\bibnamefont{Rossi}},
  \bibinfo{journal}{GNU Scientific Library Reference Manual}
  (\bibinfo{year}{2009}), \bibinfo{note}{3rd ed., Network Theory, Ltd.,
  London}.

\bibitem[{\citenamefont{Laughlin}(2006)}]{Laughlin2006}
\bibinfo{author}{\bibfnamefont{R.~B.} \bibnamefont{Laughlin}},
  \bibinfo{journal}{Philosophical Magazine} \textbf{\bibinfo{volume}{86}},
  \bibinfo{pages}{1165} (\bibinfo{year}{2006}).

\bibitem[{Note2()}]{Note2}
Note2, \bibinfo{note}{the free energy $F_0$ in minimum (for $T=0$) is equal to
  $W$ (cf. Eq.~\ref {eq:W}). If $\Delta _T^c/\Delta _S^c \equiv \Delta
  _T/\Delta _S < 10^{-4}$ then the impact of $\Delta ^c_T$ for the final energy
  of the solution is about $10^{-8}$ smaller than the impact of $\Delta ^c_S$.
  Thus $\Delta _T$ in practical calculations can be neglected.}

\bibitem[{\citenamefont{Plakida}(2010)}]{Plakida}
\bibinfo{author}{\bibfnamefont{N.}~\bibnamefont{Plakida}},
  \emph{\bibinfo{title}{High-Temperature Cuprate Superconductors: Experiment,
  Theory, and Applications}} (\bibinfo{publisher}{Springer},
  \bibinfo{address}{New York}, \bibinfo{year}{2010}).

\bibitem[{\citenamefont{Kaczmarczyk et~al.}(2013)\citenamefont{Kaczmarczyk,
  Spa\l{}ek, Schickling, and B\"unemann}}]{Kaczmarczyk2013-PRB.88.115127}
\bibinfo{author}{\bibfnamefont{J.}~\bibnamefont{Kaczmarczyk}},
  \bibinfo{author}{\bibfnamefont{J.}~\bibnamefont{Spa\l{}ek}},
  \bibinfo{author}{\bibfnamefont{T.}~\bibnamefont{Schickling}},
  \bibnamefont{and}
  \bibinfo{author}{\bibfnamefont{J.}~\bibnamefont{B\"unemann}},
  \bibinfo{journal}{Phys. Rev. B} \textbf{\bibinfo{volume}{88}},
  \bibinfo{pages}{115127} (\bibinfo{year}{2013}).

\end{thebibliography}

\end{document}